\documentclass[aps,prb,twocolumn, showpacs,superscriptaddress,amsmath,amssymb]{revtex4}
\usepackage{graphicx}
\usepackage{amsmath}
\usepackage{textcomp}
\usepackage{graphicx}
\usepackage{dcolumn}
\usepackage{bm} 
\usepackage{color}
\usepackage{ulem}

\begin{document}



\title{%
Crystal and magnetic structure of the superconductor CeNi$_{0.8}$Bi$_2$}
\author{K.~Kodama}
\affiliation{Quantum Beam Directorate, Japan Atomic Energy Agency, Tokai, Ibaraki 319-1195, Japan }
\affiliation{JST, Transformative Research-Project on Iron Pnictides (TRIP), Tokyo 102-0075, Japan }
\author{S.~Wakimoto}
\affiliation{Quantum Beam Directorate, Japan Atomic Energy Agency, Tokai, Ibaraki 319-1195, Japan }
\affiliation{JST, Transformative Research-Project on Iron Pnictides (TRIP), Tokyo 102-0075, Japan }
\author{N.~Igawa}
\affiliation{Quantum Beam Directorate, Japan Atomic Energy Agency, Tokai, Ibaraki 319-1195, Japan }
\author{S.~Shamoto}
\affiliation{Quantum Beam Directorate, Japan Atomic Energy Agency, Tokai, Ibaraki 319-1195, Japan }
\affiliation{JST, Transformative Research-Project on Iron Pnictides (TRIP), Tokyo 102-0075, Japan }
\author{H.~Mizoguchi}
\affiliation{Frontier Research Center, Tokyo Institute of Technology, Nagatsuta, Yokohama 226-8503, Japan}
\author{H.~Hosono}
\affiliation{Frontier Research Center, Tokyo Institute of Technology, Nagatsuta, Yokohama 226-8503, Japan}
\affiliation{Materials and Structures Laboratory, Tokyo Institute of Technology, Nagatsuta, Yokohama 226-8503, Japan}


\date{\today}

\begin{abstract}
We have performed powder neutron diffraction on the new superconductor, CeNi$_{0.8}$Bi$_2$ with a superconducting transtion 
temperature $T_\textrm{c}~\sim$ 4.2~K.  The structural parameters of this compound at room temperature are determined by 
Rietveld analysis.  
Below about 5 K, the clear magnetic Bragg peaks with propagation vector $q$=(0 0 0) are observed.  
The observed intensities of magnetic Bragg peaks can be explained by the magnetic structure that the two Ce moments 
in the unit cell are antiparallel along $c$ axis.  
The magnetic Bragg peaks are observed in the superconducting state, indicating the coexistence of the antiferromagnetic 
ordering and the superconductivity in this compound.  
The intensity of magnetic Bragg peak monotonously increases with decreasing temperature below $T_\textrm{N}$ 
and does not exhibit apparent anomaly at $T_\textrm{c}$, obviously different from cases of heavy fermion superconductors 
in which the magnetic ordering and the superconductivity coexist, for example, Cd-doped CeCoIn$_5$.  
These results suggest that the 4$f$ electron of the Ce atom is not coupled with the superconducting carrier, and 
the magnetic ordering is almost independent of the superconductivity in CeNi$_{0.8}$Bi$_2$.  

\end{abstract}


\maketitle

\section{Introduction}
An iron-based superconductor was first discovered in the LaFeAsO system with a so-called ZrCuSiAs-type 
structure whose superconductivity is induced by partial substitution of flourine for oxygen.\cite{kamihara}  
After this discovery, the development of new materials with an Fe layer in which the superconductivity is 
considered to appear, have been performed, resulting in the discoveries of other iron-based superconductors, 
such as the so-called 122 system,\cite{rotter} FeSe$_{1-x}$Te$_x$,\cite{hsu} LiFeAs,\cite{wang} and so on.  
On the other hand, new superconductors with a ZrCuSiAs-type structure which consist of elements other than Fe
are also explored, for example, La$_{1-x}$Sr$_x$NiAsO\cite{fang}.  
Recently, Mizoguchi and his collaborators have discovered the superconductivity in CeNi$_{0.8}$Bi$_2$ with a superconducting 
transition temperature $T_\textrm{c}~\sim$ 4.2~K.\cite{mizoguchi}  
\par
Series compounds CeMBi$_2$ and CeMSb$_2$ have ZrCuSiAs-type structures with space group of $P4/nmm$ corresponding to 
the so-called 1111 system, where M is a transition metal element such as Mn, Fe, Ni, Cu, Ag, and so on.\cite{flandorfer}  
For CeMBi$_2$, Bi1, M, Ce and Bi2 sites correspond to As, Fe, rare earth element and O sites, respectively.  
The "parent" compound of CeNi$_{0.8}$Bi$_2$, CeNiBi$_2$ is a moderately heavy fermion antiferromagnet ; 
magnetic ordering appears at about 5~K and the electronic specific heat coefficient 
$\gamma$ is relatively large (470~mJ/K$^2$ mol).\cite{onuki}   This parent compound does not exhibit a superconductivity 
although the electric conductivity is metallic down to the lowest temperature.\cite{onuki, jung} 
\par
The superconductivity is induced by the deficiency of Ni atom,\cite{mizoguchi} as the superconductivity 
in the iron-based superconductor of the so-called 1111 system is induced by oxygen-deficiency.\cite{kito}  
The resistivity and superconducting shielding signal in the magnetic field show the superconducting transition at about 4.2~K. 
At about 5~K, specific heat $C$ exhibits a jump of $\sim$4~J/K mol.  This jump may be attributed 
to the magnetic ordering of the Ce 4$f$ moment because the entropy estimated by an integration of $C/T$ below 5~K corresponds 
with $R$ln2, which is expected from the twofold degenerated ground state of the Ce 4$f$ electron.  
The clear jump of the specific heat is not observed at around $T_\textrm{c}$ because the jump caused by the superconducting  
transition is much smaller than the jump caused by the magnetic ordering, suggesting that the charge carrier which causes the 
superconductivity is not coupled with the Ce 4$f$ electron and the mass is not enhanced.\cite{mizoguchi}  
\par   
If the Ce 4$f$ electron is coupled with the superconducting carrier in the present compound, the magnetic fluctuation of 
4$f$ electron can induce the unconventional superconducting order parameter.  
In such a case, temperature dependence of the magnetic Bragg peak should have some anomaly at $T_\textrm{c}$.  
Then we have performed powder neutron diffraction measurements on CeNi$_{0.8}$Bi$_2$ in order to investigate the 
existence of the coupling between the Ce 4$f$ electron and the superconducting carrier.  
First, we present the structural parameters determined by Rietveld analysis in Sec. IIIA.  
The elecrtonic state of this compound discussed in Ref. 6 is based on the present result.  
In Sec. IIIB, we report the magnetic structure and the temperature dependence of the ordered moment of the Ce 4$f$ moment.  
Our results suggest that the Ce 4$f$ electron does not contribute to the superconductivity in this compound, consistent with 
the suggestion in ref. 6.  
\section{Experiments}
For the compound of CeNi$_x$Bi$_2$, we found through a series of synthesis experiments that samples synthesized 
by conventional solid state reaction have a constant Ni-deficiency of x=0.8.\cite{mizoguchi} 
A powder sample of CeNi$_{0.8}$Bi$_2$ for the present neutron measurements was prepared by following method.  
Ce, Ni and Bi powders with a nominal composition of CeNi$_{0.8}$Bi$_2$ were used as starting materials.  
The mixed powder was evacuated in a silica tube and heated at 500~\textcelsius~for 10 h and 750~\textcelsius~20 h.  
The obtained powder was ground and pressed into pellets, and it was heated 800~\textcelsius~for 10 h 
in the evacuated silica tube. 
Inductively coupled plasma (ICP) spectroscopy was used to confirm the chemical composition, CeNi$_{0.8}$Bi$_2$, 
which was consistent with the nominal composition.  
The sample appears stable thermally in a dry box or an evacuated glass tube, 
but decomposes in an hour when exposed to an ambient atmosphere.  
This decomposition would be induced by the reaction of an unusual valence state of Bi ions in the material 
with water vapor in the atmosphere. 
Thus, we paid attention so as to avoid decomposition or degradation of the sample.
The sample of about 14.3~g was used in neutron diffraction measurements.  
Electric resistivity and superconducting shielding measurements confirm that $T_\textrm{c}$ value and a volume 
fraction of shielding signal are the same as the values reported in ref. 6.  
\par
The powder neutron diffraction pattern for the analysis on a crystal structure was collected 
by using the high-resolution powder diffractometer (HRPD) installed in the reactor JRR-3 of Japan Atomic Energy Agency (JAEA).  
The neutron wavelength was 1.8234 $\mathrm{\AA}$ and the collimation was 
open (effective value of 35')-40'-S-6' (S denotes sample). 
The pattern was collected at room temperature.  
The powder sample was set in a vanadium holder enclosed in Al cans filled with He gas 
in order to avoid the decomposition in the atmosphere.
Diffraction data for the analysis on the magnetic structure were collected by using the triple-axes spectrometer TAS-2 
in reactor JRR-3 of JAEA.  
The neutron wavelength used was 2.3590 $\mathrm{\AA}$ and the collimation was 14'-40'-S-40'-80'. 
The sample enclosed in Al can was mounted in a closed-cycle refrigerator.

\section{Results and Discussions}
\subsection{Analysis of crystal structure}
Figure 1 shows the neutron powder diffraction pattern of CeNi$_{0.8}$Bi$_2$ obtained from the HRPD.  
\begin{figure}[tbh]
\centering
\includegraphics[width=8.5cm]{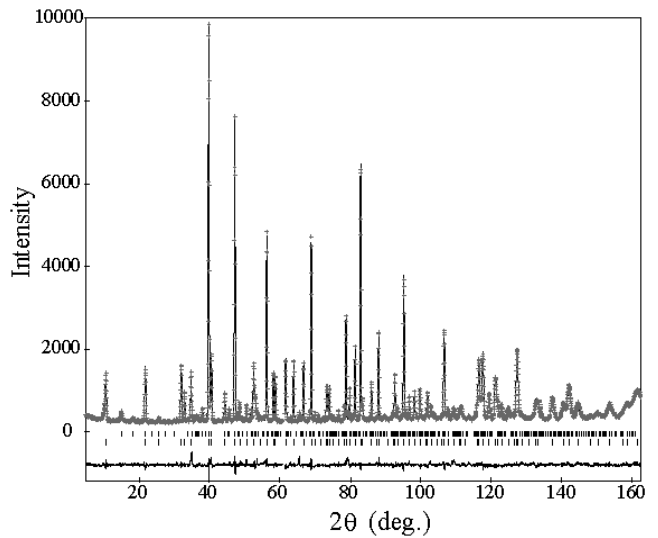} 
\caption{Observed (crosses) and calculated (solid lines) neutron powder diffraction pattern of CeNi$_{0.8}$Bi$_2$.  
The observed pattern is collected by using the HRPD at room temperature.  
Vertical bars show the calculated position of Bragg reflections including the impurities.  
The solid line at the bottom of the figure is the difference between observed and calculated intesities}
\label{fig.1}
\end{figure}
The observed data are shown by crosses.   
Structural analysis on the neutron powder diffraction pattern is performed by using the program RIETAN2000.\cite{izumi}  
The space group of $P4/nmm$ is used because we know that the series compounds of CeMBi$_2$ and CeMSb$_2$ have 
same structure as, shown in Sec. I.  
The diffraction pattern is analyzed including NiBi$_3$, NiBi and Ni as impurities.  
Because a small peak which can not be assigned by above impuries is observed at 2$\theta \sim$ 61.2~\textdegree, 
the intensity data in the 2$\theta$ range of 60.9-61.5~\textdegree~ are excluded in the analysis.  
The occupation factor of Ni is fixed at 0.8 which is determined by ICP spectroscopy.  
The obtained structural parameters of CeNi$_{0.8}$Bi$_2$ are shown in Table I.  
\begin{table}
\caption{Atomic positions of CeNi$_{0.8}$Bi$_2$ (space group $P4/nmm$) determined by Rietveld analysis 
of neutron powder diffraction data at room temperature.  Obtained lattice parameters are $a$=4.5439(1) 
and $c$=9.6414(2)~\AA. 
The $R$-factor, $R_\textrm{wp}$, is 6.04~$\%$,  }
\label{t1}
\begin{center}
\begin{tabular}{lllllll}
\hline
Atom & Site & Occ. & $x$ & $y$ & $z$ & $B$~(\AA$^{2}$) \\
\hline
Ce & 2$c$ & 1 & 1/4 & 1/4 & 0.2691(3) & 0.83(4) \\
Ni & 2$b$ & 0.8 & 3/4 & 1/4 & 1/2 & 1.83(4) \\
Bi1 & 2$c$ & 1 & 1/4 & 1/4 & 0.6384(2) & 0.92(3) \\
Bi2 & 2$a$ & 1 & 3/4 & 1/4 & 0 & 0.66(4) \\
\hline
\end{tabular}
\end{center}
\end{table}
The errors of the parameters shown in the table are mathematical standard deviations obtained by the analysis.  
The diffraction pattern calculated by using refined parameters is shown in Fig. 1 by solid lines.  
The calculated line can reproduce the observed data.  
Mass fractions of the impurities are 9.1$~\%$ for NiBi$_3$, 1.4$~\%$ for NiBi and 0.8$~\%$ for Ni, respectively.  
The larger thermal parameter of Ni relative to other sites may be due to the impurities.

\subsection{Magnetic structure and $T$ dependence of ordered Ce moment}
Results of 2$\theta$ scan at 11.2~K and 2.9~K, obtained from TAS-2 are shown in Fig. 2(a) by gray and black lines, 
respectively.  
\begin{figure}[tbh]
\centering
\includegraphics[width=7.5 cm]{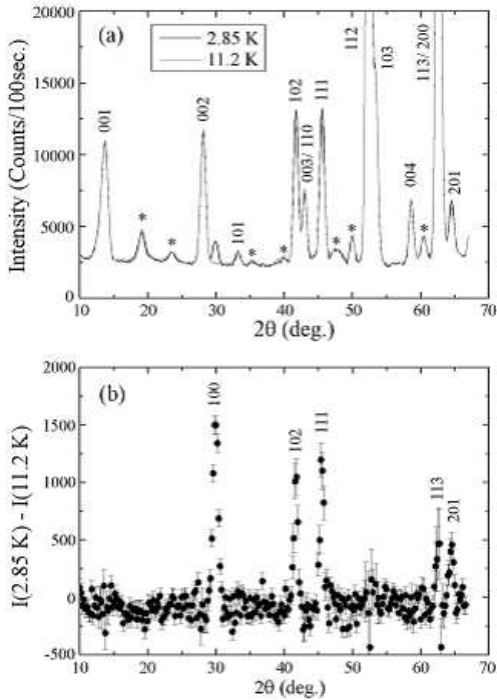} 
\caption{(a) Neutron powder diffraction patterns of CeNi$_{0.8}$Bi$_2$ obtained at 11.2~K (gray line) and 2.9~K (black line). 
The data are collected by using TAS-2.  Miller indices are shown for nuclear Bragg peaks.  
The peaks marked by asterisks are contributed from impurities, for example, NiBi$_3$. 
(b) Magnetic Bragg intensity obtained by subtracting the intensity at 11.2~K from the intensity at 2.9~K.}
\label{fig.2}
\end{figure}
At 2.9 K a clear Bragg peak appears at right-hand side of 002 reflection, corresponding to 100 reflection.  
The intensities of 102 and 111 reflections are slightly enhanced at 2.9 K.  
These results show the existence of magnetic ordering at lower temperatures and the coexistence of magnetic ordering 
and the superconductivity.  
Figure 2(b) shows the intensity obtained by subtracting the intensity at 11.2~K from the intensity at 2.9~K. 
Clear magnetic Bragg peaks are observed at lower temperatures at the reciprocal lattice points corresponding to 100, 102, 111, 
and 201.  The data are scattered around 2$\theta \sim$ 53~\textdegree~and 63~\textdegree, 
due to large nuclear Bragg intensities of 112/103 and 113/200 reflections whose intensities are about 70000 and 50000 counts 
at their peak positions, respectively. 
Magnetic 113 reflection does not have clear peak profile due to the large experimental error mentioned above, 
although it should have almost the same intensity as the magnetic 201 reflection in the magnetic structure presented below.   
\par 
From these results, we can consider the magnetic structure at low temperatures.
For 10$l$ reflections, the clear magentic Bragg peaks are observed at $l=2n$ and they are nearly absent at $l=2n+1$, 
indicating that the magnetic moments of two Ce atoms in a unit cell at (1/4 1/4 0.2691) and (3/4 3/4 0.7309) are antiparallel.  
This intesity modulation of magnetic 10$l$ reflections against $l$ also shows that Ni moments are 
almost independent of the observed magnetic Bragg peaks even if the Ni atoms have magnetic moments.  The scattering intensity 
from the Ni moments should not depend on $l$ because two Ni atoms are located on a flat layer at $z$=1/2.  
In magnetic susceptibility measurements, the contribution of Ni atom to the Curie-Weiss-like susceptibility is 
nearly absent.\cite{mizoguchi2, onuki}  
These results suggest that Ni atom is nonmagnetic.  
The clear peaks are not observed at 00$l$ positions in Fig. 2(b).   
If the Ce moments are perpendicular to $c$ axis, magnetic 00$l$ reflections with $l=2n+1$ must have large intensities. 
Especially, the intensity of 001 reflection is expected to be about four times as large as the intensity of 100 reflection.  
The absence of magnetic 00$l$ reflections indicates that the two Ce moments are oriented along $c$ axis.  
It is consistent with the anisotropy of the magnetic susceptibility, 
indicating that the antiferromagnetic easy-axis is the $c$ axis.\cite{onuki}  
Then we can get the magnetic structure shown in Fig. 3.  
\begin{figure}[tbh]
\centering
\includegraphics[width=5.5 cm]{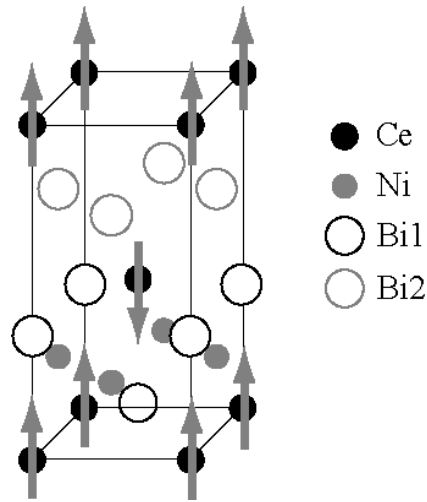} 
\caption{Crystal and magnetic structure of CeNi$_{0.8}$Bi$_2$.  The Ce atom at (1/4 1/4 0.2691) is the origin. 
Gray arrows show the magnetic moments of Ce atoms.}
\label{fig.3}
\end{figure}
In the figure, Ce atom at (1/4 1/4 0.2691) is located at the origin.
\par
The integrated intensities of observed nuclear and magnetic Bragg reflections, $I_{obs}$, are shown in Table II.  
The former intensities are estimated by using the diffraction data at 11.2 K and the latter ones are estimated 
by using the data shown in Fig. 2(b). 
\begin{table}[tbh]
\caption{Observed and calculated intensities of nuclear (left-hand side) and magnetic (right-hand side) Bragg peaks. 
Calculated intensities are obtained by using the structural parameters shown in table I.  
In the calculation of magnetic Bragg intensity, we use the amplitude of Ce moment of 1.43 $\mu_\textrm{B}$ 
and isotropic magnetic form factor of Ce$^{3+}$\cite{inter}.}
\label{t2}
\begin{center}
\begin{tabular}{llllll}
\hline
\multicolumn{6}{l}{ nuclear Bragg at 11.2 K~~~~~~~~~~~~~magnetic Bragg at 2.9~K}\\
$h k l$ & $I_{obs}$ & $I_{cal}$ & ~~~~~~~$h k l$ & $I_{obs}$ & $I_{cal}$ \\
\hline
0 0 1 & 8354(428) & 10316 & ~~~~~~~1 0 0 & 1579(128) & 1594  \\
0 0 2 & 8354(152) & 8607 & & & \\
1 0 1 & 759(93) & 652 & & &  \\
1 0 2 & 7148(301) & 7734 & ~~~~~~~1 0 2 & 827(141) & 715 \\
003/110 & 3643(200) & 3843 & & & \\
1 1 1 & 7490(304) & 5111  & ~~~~~~~1 1 1 & 905(149) & 1008 \\
1 1 2 & 51380(951) & 52171  & & & \\
1 0 3 & 10002(325) & 7569 & & &  \\
0 0 4 & 3626(108) & 3520 & & & \\
113/200 & 38757(1001) & 40004 & & & \\
2 0 1 & 2842(231) & 2418 & ~~~~~~~2 0 1 & 363(123) & 200 \\
\hline
\end{tabular}
\end{center}
\end{table}
The intensities of nuclear Bragg reflections calculated by using the structural parameters shown in table I, $I_{cal}$, 
are also shown in right-hand side of $I_{obs}$ in the table.  
The calculated intensities almost correspond with the observed intensities.  
To estimate the ordered moment of Ce sites at 2.9 K, the intesities of magnetic reflections 
obtained from the magnetic structure shown in Fig. 3 are fitted to the observed intensities 
by using least-squares fitting program.  
The estimated value of the ordered moment is 1.43(5) $\mu_B$.  
Here, we use the isotropic magnetic form factor of Ce$^{3+}$.\cite{inter}  
The calculated intensities of magnetic Bragg reflections are shown in the right-hand side of the table.  
The calculated intensities agree with the observed intensities, indicating that present magnetic structure and 
the obtained ordered moment are reasonable.  
Although 113 reflection should have the intensity of $\sim$200 in the calculation, the observed intensities can not be 
estimated due to the experimental error, as mentioned above.
\par
The inset of Fig. 4(a) is the temperature ($T$) dependence of the peak profile of 100 reflection.  
\begin{figure}[tbh]
\centering
\includegraphics[width=7.5 cm]{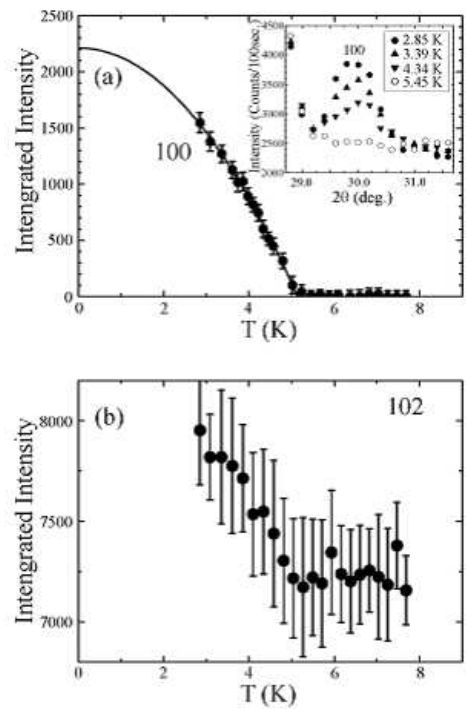} 
\caption{Temperature dependences of the integrated intensities of (a) 100 and (b) 102 reflections, respectively. 
The solid line shows the fitting the data below 5~K to the function that the intensity is proportional to 
$1-(T/T_\textrm{N})^2$ and the extrapolation of the fitting function to zero kelvin.  
In the inset : temperature dependence of the peak profile of 100 reflection.}
\label{fig.4}
\end{figure}
As shown in the inset, 100 reflection appears below 5.45~K and the peak gradually develops with decreasing $T$.  
Figures 4(a) and 4(b) show the $T$ dependences of the integrated intensities of 100 and 102 reflections, respectively.  
The intensity of 102 reflection has a large error bar in this scale because this reflection includes the nuclear scattering.  
The intensities of both reflections increase with decreasing $T$ below about 5~K, consistent with the temperature 
at which the jump of the specific heat is observed.\cite{mizoguchi}  
As shown in the main panel of Fig. 4(a), the temperature development of the magnetic Bragg intensity is monotonous 
down to 2.9~K and an anomaly is not observed at the superconducting transtion temperature, 4.2~K.  It is quite different 
from the behaviors observed in heavy fermion superconductors in which the magnetic ordering and the superconductivity coexist.  
For example, Cd-doped CeCoIn$_5$, CeCo(In$_{1-x}$Cd$_x$)$_5$ with $x$=0.1 and 0.075, exhibit the antiferromagnetic orderings 
of the Ce moment at Neel temperatures, $T_\textrm{N}~\sim$ 3.0~K and 2.5~K, respectively.  
Superconductivities appear below $\sim$ 1.2~K and 1.7~K,  
and the antiferromagnetic ordering and the superconductivity coexist below these temperatures.\cite{pham}  
In these compounds, the magnetic Bragg peaks appear below $T_\textrm{N}$ and their intensities increase with decreasing $T$.  
However, the developments of the magnetic Bragg intensities are suppressed and they become almost flat to the temperature 
below $T_\textrm{c}$.\cite{nicklas, nair}  The present $T$ dependence of the magnetic Bragg intensity suggests that 
the Ce 4$f$ electron is almost independent of the superconductivity and this compound is not a heavy fermion superconductor.  
\par
The solid line in Fig. 4(a) shows the fitting of data below 5.0~K to the function that the intensity is proportional 
to $1-(T/T_\textrm{N})^2$ and the extrapolation of the fitting function to 0 K.  The amplitude 
of the momtent at 0 K obtained by the fitting is 1.71 $\mu_B$.  
In the experimental accuracy, any structural change is not observed and the tetragonal symmery remains down to 2.9~K.  
In the tetragonal symmetry, energy levels of the Ce 4$f^1$ state split into three doublets, $\Gamma_7^{(1)}$, $\Gamma_7^{(2)}$, 
and $\Gamma_6$, by the crytalline electric field, where the two former doublets consist of linear combination of $J_z=\pm 3/2$ 
and $J_z=\pm 5/2$ states, and the latter consists of $J_z=\pm 1/2$ state.    
In the case of CeAgSb$_2$ which also has a ZrCuSiAs-type structure, the ground state is the $\Gamma_6$ state, 
revealed by the measurement of the crytalline electric-field excitation.  The amplitude of the antiferromagnetic ordered moment 
is in good agreement with $g_J \mu_B J_z \sim$ 0.43~$\mu_B$.\cite{araki}  
On the other hand, the ordered moment in this compound is much larger than the moment of CeAgSb$_2$, 
indicating that the ground state of this compound is not the $\Gamma_6$ state.  
Such differences of the amplitude of the ordered moment and/or the ground state 
are caused by the differences of the structural parameters and constituent elements.  
Actually, other compounds with a ZrCuSiAs-type structure, CeCuBi$_2$ and CeAgBi$_2$, exhibit antiferromagnetic ordering 
with easy axes parallel to the $c$ direction.  Their saturated ordered moments which are estimated by magnetization curves 
are 1.76 and 2.1 $\mu_B$, respectively.   These behaviors are similar to the present compound while in CeCuBi$_2$ 
and CeAgBi$_2$, the superconductivity is not observed and their magnetization curves show metamagnetic behavior 
which is not observed in the present compound.    
\par
CeNi$_{0.8}$Bi$_2$ has the same crystal structure as the so-called 1111 system which has a maximum $T_\textrm{c}$ of about 55~K 
among iron-based high-$T_\textrm{c}$ superconductors.  
Theoretical studies on an iron-based high-$T_\textrm{c}$ superconductor point out that 
magnetic fluctuation arising from a nesting between Fermi surfaces at $\Gamma$ and M points 
in an unfolded Brillioun zone is an origin of the superconductivity\cite{mazin,cvetkovic,kuroki,ma}.  
Experimentally, strong corelations between the high-$T_\textrm{c}$ superconductivity and the spin fluctuation are observed in 
an iron-based high-$T_\textrm{c}$ superconductor.  
The inelastic neutron scattering measurements on LaFeAsO$_{1-x}$F$_x$ with $x$=0.057, 0.082 and 0.157, show that 
the spin fluctuation observed in the samples with $x$=0.057 and 0.082, whose $T_\textrm{c}$ are about 30~K, almost disappears 
in the sample with $x$=0.157 in which the high-$T_\textrm{c}$ superconductivity is almost suppressed.\cite{wakimoto}  
On the other hand, the superconductivity may appear without magnetic fluctuation in CeNi$_{0.8}$Bi$_2$, 
because Ni is nonmagnetic and the Ce moment is almost decoupled with the superconducting carrier.  
TheaAbsence of the magnetic fluctuation may indicate that the $T_\textrm{c}$ of CeNi$_{0.8}$Bi$_2$ is much lower than 
the values of iron-based superconductors although they have same crystal structures.  

\par
\section{Summary}
We have performed neutron powder diffraction measurements on a new superconductor, CeNi$_{0.8}$Bi$_2$, with  
$T_\textrm{c}~\sim$ 4.2~K. 
The structural parameters at room temperature are determined by the Rietveld method.  
The clear magnetic Bragg peaks are observed at $q$=(0 0 0) below about 5~K.  
In the magnetic ordering phase, the two Ce moments in the unit cell are antiparallel along the $c$ axis and   
the Ni atom does not contribute to the magnetic Bragg reflections.  
Below $T_\textrm{c}$, the superconductivity and antiferromagnetic ordering coexist.  
The magnetic Bragg intensity monotonously increases with decreasing $T$ below about 5~K and does not exbihit an apparent 
anomaly at $T_\textrm{c}~\sim$ 4.2~K, suggesting that the 4$f$ electron of the Ce atom is almost independent of the 
superconductivity.  The saturated ordered moment is about 1.7 $\mu_\textrm{B}$.  
\par
\begin{acknowledgments}
The authors would like to thank K. Kaneko for his fruitful discussion.  
This work was supported by the Funding Promgram for World-Leading Innovative R\&D on Science and Technology (FIRST), Japan,  
and also supported by a Grant-in-Aid for Specially Promoted Research 17001001 from 
the Ministry of Education, Culture, Sports, Science and Technology, Japan.
\end{acknowledgments}


\begin{thebibliography}{9}
\bibitem{mizoguchi2}
Magnetic susceptibility of CeNi$_{0.8}$Bi$_2$ can be fitted by the Curie-Weiss law in the temperature range 
of 30 $\le T \le$ 300~K.  
From the Curie constant, the effective moment is 2.80 $\mu_B$, which corresponds to the effective moment of single crystal 
CeNiBi$_2$, 2.88 $\mu_B$ and 2.60 $\mu_B$ for $H \parallel a$ and $H \parallel c$, reported in ref. 8.  
These values are larger than 2.54 $\mu_B$ of a free Ce$^{3+}$ ion.  
However, if the Ni atom has a local moment and also contributes to the Curie-Weiss-type susceptibility, 
the Curie constant and/or the effective moment have to be much larger than the observed values.
\end{thebibliography}
\end{document}